\documentclass[aps,showpacs,showkeys,12pt,preprintnumbers,nofootinbib]{revtex4}
\usepackage{graphicx}
\usepackage{color}
\usepackage{amsmath}
\usepackage{amssymb}
\usepackage{multirow}
\usepackage{lineno}
\usepackage{booktabs}
\usepackage{rotating,tabularx}
\usepackage{tabu}
\usepackage{lscape}

\makeatletter
\renewcommand\paragraph{\@startsection{paragraph}{4}{\z@}%
            {-2.5ex\@plus -1ex \@minus -.25ex}%
            {1.25ex \@plus .25ex}%
            {\normalfont\normalsize\bfseries}}
\makeatother
\def \be  {\begin{equation}}
\def \ee  {\end{equation}}
\def \ee  {\end{equation}}
\def \bea {\begin{eqnarray}}
\def \eea {\end{eqnarray}}

\begin{document}
\hspace{0.05cm}


\title{Analyzing the Correlation Between Thermal and Kinematic Parameters in Various Multiplicity Classes within 7 and 13 TeV pp Collisions}

\author{Muhammad Waqas}
\email{20220073@huat.edu.cn, waqas_phy313@yahoo.com}
\affiliation{School of Mathematics, Physics and Optoelectronic Engineering, Hubei University of Automotive Technology 442002, Shiyan, People's Republic of China}

\author{Wolfgang Bietenholz}
\email{wolbi@nucleares.unam.mx}
\affiliation{Instituto de Ciencias Nucleares - Universidad Nacional Autónoma de México,  Apartado Postal 70-543, CdMx 04510, Mexico} 

\author{Mohamed Bouzidi}
\affiliation{Department of Physics, Faculty of Science, University of Ha'il, P.O.Box 2440, Ha'il, Saudi Arabia} 

\author{Muhammad Ajaz}
\email{ajaz@awkum.edu.pk}
\affiliation{Department of Physics, Abdul Wali Khan University Mardan, 23200 Mardan, Pakistan}

\author{Abd Al Karim Haj Ismail}
\affiliation{College of Humanities and Sciences, Ajman University, PO Box 346, UAE}

\author{Taoufik Saidani}
\affiliation{Department of Computer Sciences, Faculty of Computing and Information Technology, Northern Border University, Rafha, Saudi Arabia}

\begin{abstract}
We investigate the transverse momentum spectra of identified particles at 7 TeV and 13 TeV in pp collisions in the framework of the blast wave model with Tsallis statistics (TBW). Based on experimental data by ALICE Collaboration, we observe that the model describes the $p_T$ spectra well with the common Tsallis temperature ($T$) and flow velocity ($\beta_T$) but separate non-extensive parameters ($q$) for baryons and mesons. The parameter dependence on multiplicity as well as on collision energy is investigated, and a strong dependence on the former while a weak dependence on the latter is reported. The extracted parameters in this work consist of the initial temperature (\(T_i\)), the average transverse momentum (\( \langle p_T \rangle \)), the Tsallis temperature (\(T\)), flow velocity (\(\beta_T\)), and the non-extensive parameter (\(q\)). These parameters are found to increase a little with increasing energy, however, they (except the parameter $q$) decrease significantly with decreasing multiplicity. We observe that $\beta_T$ drops to zero after the multiplicity class VII, while, $T$ and $q$ do not change their behavior. 
Furthermore, our analysis explore the correlations among different parameters, including associations with the charged particle multiplicity per unit pseudorapidity (\(\langle dN_{ch}/d\eta\rangle\)). The correlation between \(T\) and \(\beta_T\),
$T$ and \( \langle dN_{ch}/d\eta\rangle \), $\beta_T$ and \( \langle dN_{ch}/d\eta\rangle \), $T_i$ and \( \langle p_T \rangle \) and $T_i$ and \( \langle dN_{ch}/d\eta\rangle \)  demonstrates a positive relationship, 
while, the correlation between \(T\) and \(q-1\), and $q-1$ and \( \langle dN_{ch}/d\eta\rangle \) is negative.  
Finally, we implement an extra flow correction on the \(T\) parameter. Our findings reveal that the Doppler-corrected temperature parameter aligns closely with the \(T\) in scenarios with lower multiplicities. However, as the multiplicity increases, a noticeable divergence emerges between these parameters, indicating a widening separation between them.

\end{abstract}


\date{\today}

\maketitle

\section{Introduction}
The spectra of transverse momenta (\( p_T \)) of particles in their final state within high-energy collisions serve as crucial probes into the system dynamics. This analytical approach proves particularly effective in acquiring insights into the Quark Gluon Plasma (QGP), renowned for its hot and dense characteristics. Within the domain where \( p_T \leq 2 \) GeV/c, hadrons originate from soft scattering events. Analyzing their spectra provides crucial data about fundamental system properties such as the kinetic freezeout temperature (\( T_0 \)) and \( \beta_T \). Extracting these attributes heavily relies on hydrodynamic modeling, often incorporating the Boltzmann-Gibbs blast-wave (BGBW) model \cite{Schnedermann:1993ws}, as discussed in \cite{DerradideSouza:2015kpt}. Conversely, in regions where \( p_T \geq 2 \) GeV/c, hadrons result from hard scatterings involving partons. Their properties are interpreted within perturbative Quantum Chromodynamics (pQCD).
Previously, observations from pp collisions at the Relativistic Heavy-ion Collider (RHIC) and the Large Hadron Collider (LHC) served as benchmarks for understanding particle production in heavy-ion collisions. However, recent observations of heavy-ion-like behavior in high-multiplicity pp collisions at \( \sqrt{s} = 7 \) TeV raise doubts about using pp collisions at LHC energies as a sole baseline for heavy-ion studies \cite{1,2}.

Proton-proton collisions with high multiplicity can produce an increased numbers of final state hadrons, potentially favoring system thermalization through momentum transfer in multiple partonic interactions. Such complex systems can be described by statistical models with a few parameters. The study of transverse momentum spectra informs us about kinetic freeze-out processes in these high-energy collisions. The application of Tsallis non-extensive statistics \cite{15} introduces a power-law distribution, particularly influential at higher collision energies, offering a better fit to the transverse momentum spectra of various particles \cite{3,4,5,6,7,8,9,10,11,12,13,14,15,16}. This approach accounts for temperature fluctuations, either on an event-by-event basis or within a single event \cite{17}. These fluctuations are linked to the non-extensive parameter \( q \) \cite{18}, indicating the system's deviation from equilibrium. A recent introduction of $q$-dual entropy provides a thermodynamic link to the Tsallis distribution \cite{19}.

Collisions involving hadrons traveling at relativistic speeds provide an opportunity to explore the properties of dense nuclear matter at temperatures significantly surpassing the pseudo-critical QCD temperature \(T_c = 156.5 \pm 1.5\) MeV \cite{21}. These conditions create an environment conducive to unbound quarks and gluons.
The substantial production of hadrons and light nuclei strongly suggests that thermal particle generation occurs at a constant chemical freeze-out temperature \(T_{\text{chem}} \approx T_c\). This observation is supported by the statistical hadronization model (SHM) \cite{22}, affirming the presence of a consistent chemical freeze-out temperature akin to the pseudo-critical QCD temperature. Additionally, models employing a viscous fluid depiction of QGP expansion accurately capture various soft hadronic observables \cite{23,24,26,27}. Comprehensive fits to empirical data enable the extraction of model parameters and the dense QCD matter's transport characteristics \cite{28}.
The blast-wave model, one of the earliest and most straightforward frameworks, describes hadron production in a dynamic medium \cite{Schnedermann:1993ws,29}. This approach utilizes a parameterized freeze-out surface characterized by  \(T\) and \(\beta_T\). It assumes that the primary particle spectra conform to a thermal distribution within the fluid's local rest frame. Experimental observations of hadrons, such as charged pions, kaons, and protons, are computed by combining decay contributions from short-lived primary resonances with the initial thermal abundances. 
While direct decay freeze-out often provides a satisfactory fit to the data \cite{30,31,32}, it overlooks potential hadron re-scattering and regeneration processes, which can be accounted for through the inclusion of hadronic after-burners \cite{33,34}. An alternative simplification involves fitting the thermal spectra of pions, kaons, and protons directly to the observed particle spectra. This approach incorporates individual normalization and constrains momentum ranges for the fit \cite{35}.
This standard analysis method is commonly employed to scrutinize and compare soft particle production across various collision systems and centralities \cite{35,36,37,38}.

This study unveils a fitting methodology based on the blast-wave model, which integrates the Tsallis distribution. Traditional Boltzmann-Gibbs blast-wave (BGBW) modeling presupposes a medium with uniform \( \beta_T \) that experiences an abrupt kinetic freeze-out at some temperature \( T_0 \). However, given the variances in initial states for the hydrodynamic development in AA and pp collisions, these assumptions don't entirely hold. Fluctuations in both scenarios imprint themselves on particle spectra in the low to intermediate \( p_T \) domain \cite{39}.
Such conditions are addressed by introducing non-extensive Tsallis statistics into the BGBW model, creating what is termed the Tsallis blast-wave model. This hybrid model has been effective in describing particle spectra in different collision types at LHC and RHIC energies \cite{41, 41a}. An extension of this model to pp collisions has even identified a radial flow onset beyond 0.9 TeV collision energy \cite{42}.
The current investigation employs the TBW to dissect the \( p_T \) spectra of identified particles in various multiplicity classes at 7 and 13 TeV. We aim to reveal how $\beta_T$, \(T\), and the non-equilibrium parameter \( q \) vary depending on both collision energy and multiplicity class. Identified particles were specifically studied as they serve as reliable markers for particle production dynamics. Furthermore, different correlations are studied which provide useful information about the properties of the QGP, including the equation of state (EOS), temperature evolution, flow patterns, and nature of particle production.

The paper's structure is delineated as follows: 
Section II provides a succinct overview of the thermodynamically consistent Tsallis distribution and the BGBW model, which are employed to describe the \(p_T\)-spectra of charged particles at \( \sqrt{s} = 7 \) and 13 TeV.
Section III delves into the outcomes derived from utilizing Tsallis non-extensive statistics and the BGBW model. It scrutinizes these outcomes as they relate to charged particle multiplicity and collision energy.
Lastly, Section IV encapsulates and summarizes the findings obtained from the study.

\section{Formalism and method}

We delve into the intricate nature of high-energy collisions within the theoretical structure of a blast-wave model augmented by Tsallis statistics \cite{42}. A plethora of distributions, encompassing Erlang \cite{52}, canonical (Boltzmann, Fermi–Dirac, and Bose–Einstein) \cite{53}, Tsallis \cite{54}, hybrid Tsallis-canonical \cite{57,61,62}, Schwinger mechanisms \cite{63,64}, and blast-wave models rooted in Boltzmann statistics \cite{36,65,68}, serve as viable candidates for depicting the multifaceted features of \( p_T \) spectra in soft excitation processes. Despite the diversity in potential probability density functions, none suffice to fully characterize \( p_T \) spectra, especially when extending up to 100 GeV/c in LHC collisions \cite{69}.

In practice, distinct \( p_T \) regimes have been identified, as outlined, defined as \( p_T < 4–6 \) GeV/c, \( 4–6 \) GeV/c \( < p_T < 17–20 \) GeV/c, and \( p_T > 17–20 \) GeV/c, corresponding to unique underlying mechanisms. According to \cite{70}, the diverse spectral attributes stemming from parton fragmentation and hadronization processes are reflected in these distinct \( p_T \) regimes. While medium effects are predominant in the first regime, they are subdued in the second and almost negligible in the third. The second regime is particularly noteworthy for the fusion and collective behavior of partons and strings. These strings represent the gluon-mediated force that carries quark-quark interactions. As such, the fusion and collective behavior of partons and strings are especially interesting in the second $p_T$ region. It suggests that both the fusion and collective behavior of partons (quarks and gluons) and the extended structures created by the gluon strings or color flux tubes are important in this particular region. 

We underscore the simultaneous presence of unique and universal traits in high-energy collision systems \cite{71,76,77,78}. This duality implies that a single fitting function could, in principle, model the spectra across a broad \( p_T \) range with consistent parameter values. Indeed, the Tsallis-like distribution has demonstrated this capability, fitting ATLAS and CMS spectra across a 14-orders of magnitude with invariant parameters, as reported in \cite{79}.

The Tsallis function, along with its different forms, has a broader applicability, even encompassing two- or three-component canonical distributions \cite{80}. This leads us to employ a composite model involving Tsallis statistics and a blast-wave fit for the extraction of \( T \) and \( \beta_T \) \cite{39,42}.

The probability density function for \( p_T \) within this composite model is expressed as 

\begin{align}
f_1(p_T)=&\frac{1}{N}\frac{\mathrm{d}N}{\mathrm{d}p_\mathrm{T}}= C p_T m_T \int_{-\pi}^\pi d\phi\int_0^R rdr \nonumber\\
& \times\bigg\{{1+\frac{q-1}{T}} \bigg[m_T  \cosh(\rho)-p_T \sinh(\rho) \nonumber\\
& \times\cos(\phi)\bigg]\bigg\}^\frac{-q}{(q-1)}
\end{align}

In this equation, \( N \) denotes the total number of particles, and \( C \) signifies the fit constant which normalizes the integral of $f_1(p_T)$ to unity. The term \( m_T \), or the ``transverse mass", is mathematically formulated as \( m_T = \sqrt{p_T^2 + m_0^2} \) \( =m_0 \sqrt{1 + (p_T/m_0)^2} \), which is roughly constant and equivalent to $m_0$ if $p_T$ $\ll$ $m_0$. Therefore, the additional factor, $m_T$, is only insignificant for the fits in the range of low transverse momentum, $p_T$ $\ll$ $m_0$. The parameter \( m_0 \) used above, is the (rest) mass of the hadron. Free parameters in the fit include \( T \), \( \beta_T \), and \( q \), where $T$ is the Tsallis temperature, $\beta_T$ is the transverse flow velocity and $q$ represents the non-extensive parameter, characterizing the degree of deviation of the system from equilibrium. As we know, the Boltzmann-Gibbs distributions with various temperatures are superimposed to form the Tsallis distribution. When $q$ deviates from unity, these temperatures undergo a fluctuation, however, the mean value of their reciprocals represent $1/T$. When $q$ approaches 1, the distribution tends to be the traditional blast wave model. The azimuthal angle \( \phi \), the maximum radial coordinate \( R \), and the radial coordinate \( r \) further characterize the system. The boost angle \( \rho \) is defined as \( \rho = \tanh^{-1} [\beta(r)] \). Here, \( \beta(r) \) represents a self-similar flow profile and is expressed as \( \beta(r) = \beta_S(r/R)^{n_0} \).
The index \( n_0 \) could potentially be 1, 2, or a free parameter \cite{Schnedermann:1993ws,pppp, 36, ALICE:2012pjb}. In this study, we opt for \( n_0 = 1 \), a choice that offers reduced uncertainties compared to the treatment of \( n_0 \) as a free parameter. \( \beta_S \) stands for the flow velocity at the surface of the emission region.

The Tsallis-blast-wave model can cover the narrow as well as wide $p_T$ range, where narrow $p_T$ corresponds to the soft excitation mechanisms and wide $p_T$ corresponds to the hard scattering. The hard scattering process can be described by the Hagedorn function \cite{83,84} or an inverse power law \cite{85,86,87} given by 

\begin{align}
f_H(p_T) = \frac{1}{N} \frac{\mathrm{d}N}{\mathrm{d}p_T} = A_0 p_T \left( 1 + \frac{p_T}{p} \right)^{-n_0}.
\end{align}
In this equation, \( A_0 \) serves as the normalization constant, ensuring that the integral of \( f_H(p_T) \) is unity. The terms \( p \) and \( n_0 \) are treated as free parameters in the fit. 

To accurately model the \( p_T \) spectra across an extensive range, two methodologies can be adopted, they involve the superposition of Eqs. (1) and (2).
The first approach uses a weighted sum, defined as:
\begin{align}
f_0(p_T) = k f_S(p_T) + (1 - k) f_H(p_T).
\end{align}
In this equation, \( k \) is a fraction that quantifies the contribution of the soft excitation process, \( f_S(p_T) \), while \( (1 - k) \) quantifies the fraction of the hard interaction process, \( f_H(p_T) \).

The second approach segregates the \( p_T \) spectrum into two regions, defined by the threshold \( p_1 \), as 
\begin{align}
    f_0(p_T) = A_1 \theta(p_1 - p_T) f_S(p_T) + A_2 \theta(p_T - p_1) f_H(p_T).
\end{align}

In the first approach (3), the parameter \( k \) and its complement \( (1 - k) \) act as the fractional contributions of the soft interactions and hard collisions processes, respectively. It's worth noting that Eq. (3) is inherently normalized, meaning the integral over the entire \( p_T \) range will yield 1.
On the other hand, in the second approach described in \cite{83}, \( A_1 \) and \( A_2 \) function as normalization constants. These constants are chosen such that the contributions from \( f_S(p_T) \) and \( f_H(p_T) \) are equal at the transition point \( p_T = p_1 \).

Both methodologies offer the flexibility to adapt to different \( p_T \) regimes, thereby providing a comprehensive framework for spectral analysis in high-energy collisions.
These two methods can be superposed in either of the two ways, as described by Eqs. (3) and (4).

To delve further into the analysis, the mean transverse momentum \( \langle p_T \rangle \) can be calculated using the following integral,

\begin{align}
\langle p_T \rangle = \int_0^\infty p_T f(p_T) \, dp_T \, .
\end{align}

Additionally, the initial temperature parameter \( T_i \) can be deduced through the string percolation method, as elaborated in the literature \cite{abc, abcd, acc}. It is defined as

\begin{align}
T_i = \sqrt{\frac{\langle p^2_T \rangle}{2F(\xi)}}.
\end{align}

In this equation, \( \langle p^2_T \rangle \) is given by \( \int_0^\infty p_T^2 f(p_T) \, dp_T \), and \( F(\xi) \) represents the color suppression factor.

%
\section{Results and discussions}

\begin{figure}
\begin{center}
\includegraphics[width=18.cm]{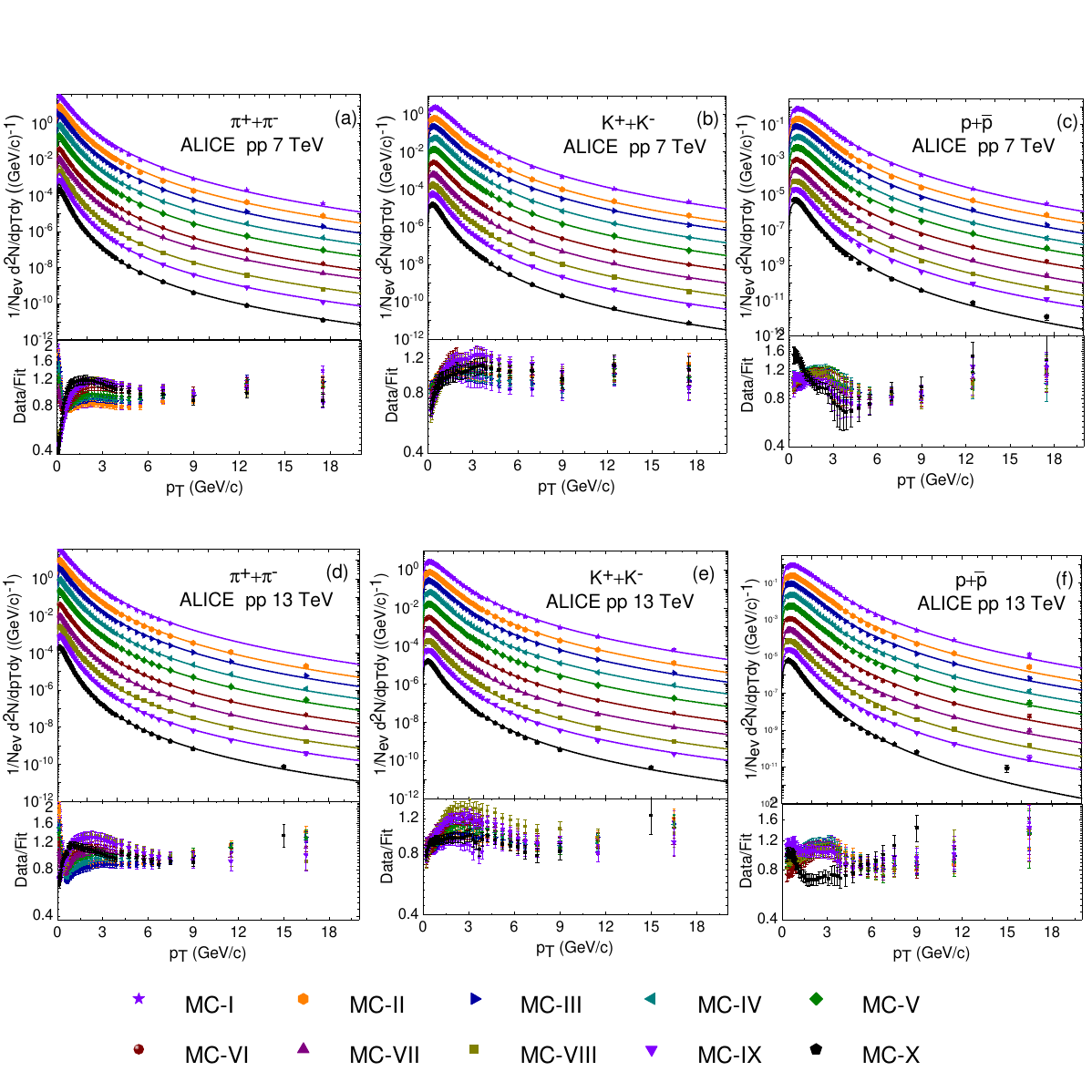}
\caption{Comprehensive analysis of the \(p_T\) spectra of \( \pi^{\pm} \), \( K^{\pm} \), and \( p (\bar{p}) \) in \( pp \) collisions at \( \sqrt{s} = 7 \) TeV and \( 13 \) TeV, divided in multiple multiplicity classes. Figures 1(a), 1(b), and 1(c) focus on the fit procedure applied to \( \pi^+ + \pi^- \), \( K^+ + K^- \), and \( p + \bar{p} \) respectively in \( pp \) collisions at 7 TeV, whereas Fig. 1(d), 1(e) and 1(f) shows the same result but for 13 TeV. The symbols are the experimental data of ALICE Collaboration \cite{ALICE:2018pal, ALICE:2020nkc}. The solid lines overlaid on the data points indicate the fit results obtained using the TBW. The fit lines affirm overlaid on the data points show the quality of the fits, indicating a good agreement between the data and the model predictions. The lower panels display the ratios of observed data to fit the results.}
\end{center}
\end{figure}

In this section, we examine the freezeout parameters associated with positively and negatively charged pions (\( \pi^{\pm} \)), positively and negatively charged kaons (\( K^{\pm} \)), and protons (anti-proton) (\( p (\bar{p}) \)) in proton-proton (\( pp \)) collisions at energies of \( \sqrt{s} = 7 \) TeV and \( 13 \) TeV. The analysis employs a sophisticated theoretical framework that melds the TBW to investigate the \( p_T \) spectra across varied multiplicity classes. The particles are divided into ten multiplicity classes starting from MC-I to MC-X where a lower multiplicity class corresponds to higher multiplicity, and higher classes correspond to lower multiplicity. 
Figure 1 serves as a graphical exposition of the fitting results. Each sub-plot is dedicated to one of the particle species under consideration, and portrays multiple multiplicity classes, as indicated by distinct markers. Specifically, Figures 1(a), 1(b), and 1(c) depict the fits for \( \pi^+ + \pi^- \), \( K^+ + K^- \), and \( p + \bar{p} \), respectively, in \( pp \) collisions at 7 TeV. In addition, Figures 1(d), 1(e), and 1(f) illustrate the corresponding fits for these particles in \( pp \) collisions at 13 TeV. 
The solid lines overlaying the data points represent the outcomes of the model fits for each multiplicity class. The extracted fitting parameters \( T \), \( \beta_T \),  and \( q \), are duly cataloged in Table 1. Additionally, statistical metrics such as \( \chi^2 \) values and the number of degrees of freedom are also enumerated in the table. The corresponding \(\chi^2\)/dof values affirm the quality of the fits, indicating a good agreement between the data and the model predictions. To prevent overlap and improve visibility, the spectra are scaled by a factor, and Table 1 lists the scaling factor. Each panel's lower half is followed by the corresponding data/fit ratios. The data/ratio in the range of $0.5$ to $1.5$ is favorable, however above 1.5 or below 0.5, indicate a not very good fit. We see that the data/fit ratio for pions at both energies are above 1.5 in the very narrow $p_T$ range ($p_T$ $\leq$0.5), because the resonance decays have a significant feed-down contribution in the corresponding $p_T$ region. 

\begin{figure}
\begin{center}
\includegraphics[width=15.cm]{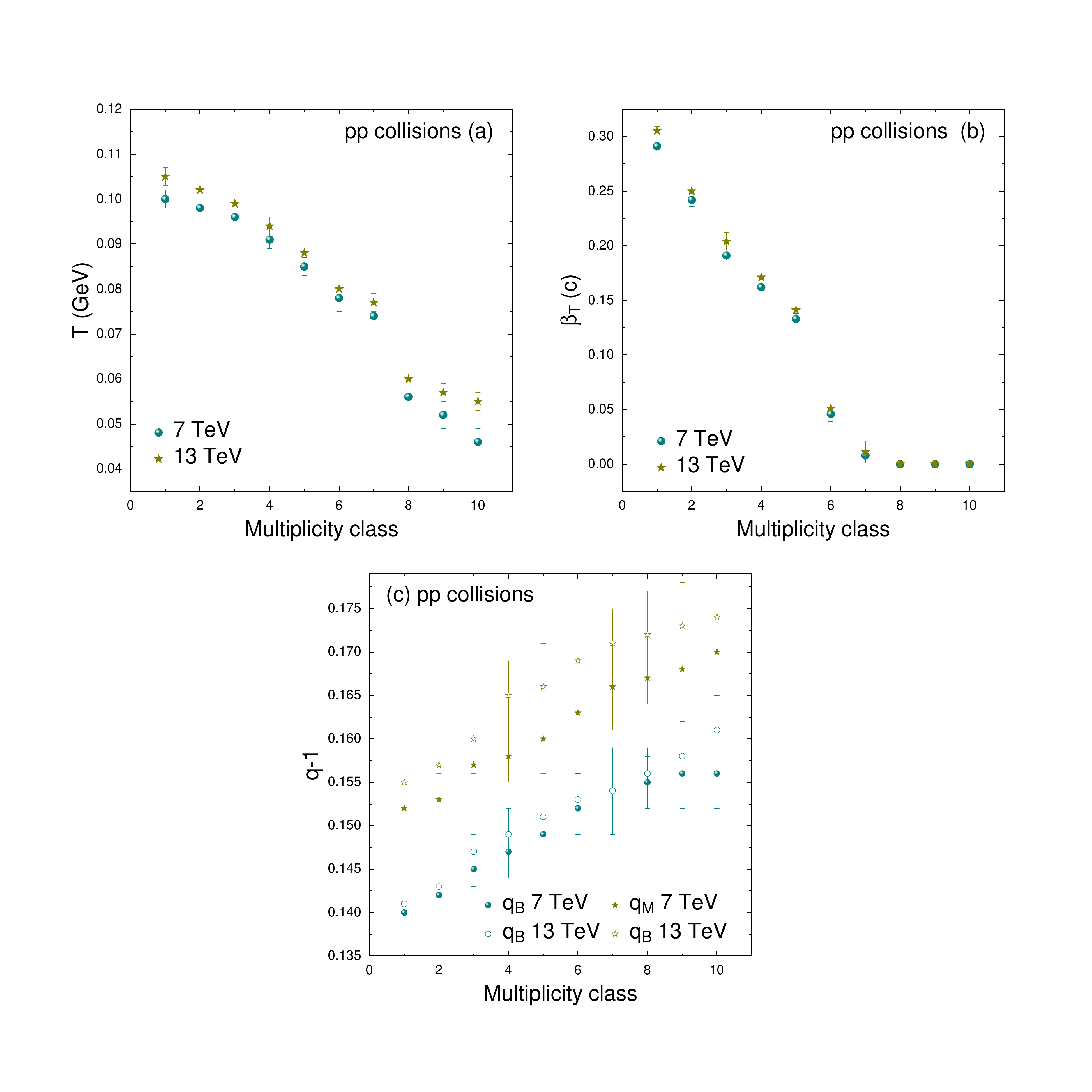}
\caption{Variation of the parameters with the multiplicity class in \( pp \) collisions at \( \sqrt{s} = 7 \) TeV and \( 13 \) TeV are shown. (a) Shows the inverse relationship between \( T \) and multiplicity class, (b) shows the decline of \( \beta_T \) with increasing multiplicity class, and (c) demonstrates the increasing trend of the non-extensive parameter $q$ (\( q_B-1 \) and \( q_M-1 \) is the non-extensivity of baryons and mesons, respectively) with the multiplicity class.}
\end{center}
\end{figure}

\begin{table*}
{\scriptsize Table 1. The values of different parameters along with $\chi^2$/dof
extracted from Fig. 1. \vspace{-.50cm}
\begin{center}
\begin{tabular}{ccccccccccc}\\ \hline\hline
  Energy &    Multiplicity Class & Scaled & $T$ (GeV)       & $\beta_T$ (c)    & $q_B-1$             & $q_M-1$       &       $\chi^2$/ dof \\ \hline
 7 TeV &  MC-I  &     & $0.100\pm0.002$ &  
  $0.291\pm0.005$  &$0.140\pm0.003$  & $0.152\pm0.03$          & 92/45\\
  -- &  MC-II  & 1/3     & $0.098\pm0.002$ &  
  $0.242\pm0.006$  &$0.142\pm0.003$  & $0.153\pm0.004$          & 88/45\\
  -- & MC-III & 1/7      & $0.096\pm0.002$ &  
  $0.191\pm0.004$  &$0.145\pm0.004$  & $0.157\pm0.004$        & 115/45\\
  -- &  MC-IV & 1/25      & $0.091\pm0.002$ &  
  $0.162\pm0.002$  &$0.147\pm0.003$  & $0.158\pm0.005$   & 141/45\\
  -- &  MC-V & 1/90      & $0.085\pm0.003$ &  
  $0.133\pm0.005$  &$0.149\pm0.004$  & $0.160\pm0.003$          & 96/45\\
  -- &  MC-VI & 1/400       & $0.077\pm0.003$ &  
  $0.046\pm0.007$  &$0.152\pm0.004$  & $0.163\pm0.004$       & 107/45\\
  -- &  MC-VII & 1400      & $0.074\pm0.002$ &  
  $0.008\pm0.0004$  &$0.154\pm0.005$  & $0.166\pm0.006$        & 46/45\\
  -- & MC-VIII & 4400      & $0.056\pm0.004$ &  
  $0$  &$0.155\pm0.003$  & $0.167\pm0.003$        & 129/45\\
  -- &  MC-IX  & 10000     & $0.052\pm0.002$ &  
  $0$  &$0.156\pm0.004$  & $0.168\pm0.004$        & 111/45\\
  -- & MC-X & 1/25000      & $0.046\pm0.004$ &  
  $0$  &$0.156\pm0.004$  & $0.170\pm0.005$      & 28/45\\
\hline
 Energy &   Multiplicity Class & Scaled  & $T$ (GeV)       & $\beta_T$ (c)    & $q_B-1$             & $q_M-1$           & $\chi^2$/ dof \\ \hline
 13 TeV &  MC-I   &    & $0.105\pm0.003$ & 
  $0.305\pm0.008$  &$0.141\pm0.003$  & $0.155\pm0.004$           & 79/48\\
  -- &  MC-II & 1/3      & $0.102\pm0.002$ &  
  $0.250\pm0.009$  &$0.143\pm0.002$  & $0.157\pm0.004$      & 94/48\\
  -- & MC-III & 1/7      & $0.099\pm0.003$ &  
  $0.204\pm0.008$  &$0.147\pm0.004$  & $0.160\pm0.004$    & 132/48\\
  -- &  MC-IV & 1/25      & $0.094\pm0.002$ &  
  $0.171\pm0.009$  &$0.149\pm0.003$  & $0.165\pm0.004$    & 140/48\\
  -- & MC-V & 1/90      & $0.088\pm0.002$ &  
  $0.141\pm0.007$  &$0.151\pm0.004$  & $0.166\pm0.005$   & 129/48\\
  -- &  MC-VI & 1/400      & $0.080\pm0.002$ &  
  $0.051\pm0.009$  &$0.153\pm0.004$  & $0.169\pm0.003$       & 45/48\\
  -- &  MC-VII & 1/1400      & $0.077\pm0.002$ &  
  $0.011\pm0.010$  &$0.154\pm0.005$  & $0.171\pm0.004$       & 53/48\\
  -- &  MC-VIII & 1/4400      & $0.060\pm0.002$ &  
  $0$  &$0.156\pm0.003$  & $0.172\pm0.005$   & 145/48\\
  -- &  MC-IX & 1/10000      & $0.057\pm0.003$ &  
  $0$  &$0.158\pm0.004$  & $0.173\pm0.005$        & 78/47\\
  -- &  MC-X & 1/25000      & $0.055\pm0.004$ &  
  $0$  &$0.161\pm0.004$  & $0.174\pm0.005$      & 157/45\\
\hline
\end{tabular}%
\end{center}}
\end{table*}
Figure 2 illustrates how the extracted parameters vary across different multiplicity classes at both 7 and 13 TeV, with each energy represented by a distinct symbol with a distinct color. Specifically, Figs. 2(a), 2(b) and 2(c) plot the \(T\), \(\beta_T\), and $q-1$ respectively, against multiplicity. The symbols horizontally from left to right demonstrate the behavior of the parameters with changing multiplicity. The values of $T$ and $\beta_T$ are the same for the particles at a given multiplicity class at a given energy, therefore we put a single symbol to represent all the particles. Notably, an inverse relationship exists between \(T\) and the multiplicity class: as the multiplicity class increases, \(T\) decreases for both energies. Additionally, for a given multiplicity class, the \(T\) values are slightly higher at 13 TeV compared to 7 TeV.

The higher \( T \) values for ``lower multiplicity class" (higher multiplicity) could indicate better thermalization. Higher multiplicity may suggest that the system reached a better thermal equilibrium, which results in higher \( T \). Similarly, higher multiplicity may lead to enhanced collective flow, contributing to higher temperatures as is indicated in the Fig. 2(b). $T$ at 13 TeV is slightly larger than that of 7 TeV, showing that the system at higher energies is slightly more excited.

Figure 2(b) reveals an inverse relationship between \( \beta_T \) and multiplicity class, with higher \( \beta_T \) values observed for lower multiplicity classes, which correspond to high-multiplicity events. A notable feature is an abrupt decline in \( \beta_T \) as one moves to higher multiplicity classes, reaching a value of zero for multiplicity classes greater than 7. The values approaching zero in these specific multiplicity classes do not imply that the fit stopped at the boundary of zero, but it indicates a significant change in the system's behavior. This trend is consistent across both 7 and 13 TeV energies. However, for multiplicity classes below 7, the \( \beta_T \) values are slightly elevated at 13 TeV compared to 7 TeV.
The sharp decrease and eventual zeroing of \( \beta_T \) may signify a shift from a regime dominated by collective effects, such as hydrodynamic flow, to a scenario where other factors become influential. This transition could be attributed to various changes in the system's energy density, the dominance of different particle production mechanisms, or alteration in the collision dynamics. The slightly larger \( \beta_T \) values at 13 TeV for the same multiplicity class could be due a to minor increase of the energy density in the system in comparison to the 7 TeV as mentioned above, and indicates that the system at 13 TeV is a little more explosive than at 7 TeV which means that the system experiences a more rapid expansion at 13 TeV. The slightly larger $T$ at 13 TeV compared to 7 TeV exhibits higher excitation of the system at 13 TeV, which is compatible with Refs. \cite{Waqas:2019mjp, Zhang:2016tbf}, but the dependence of $T$ and $\beta_T$ in this work is weaker. On the other hand, Refs. \cite{Waqas:2019mjp, Zhang:2016tbf} have analyzed $AA$ collisions where these parameters strongly depend on the collision energy. The value of $\beta_T$ at 13 TeV approaches $0.305\pm0.008$, which is close ($0.293\pm0.012$) to the peripheral (60--80\%) lead-lead (Pb-Pb) case at 2.76 TeV \cite{Che:2020fbz}. It is noteworthy that the current work focuses on pp collisions, its findings hold relevance for understanding the nature of the interacting medium in heavy-ion collision experiments. The study's investigation into freeze-out parameters, such as $T$, $\beta_T$, and $q-1$, offers insights into the characteristics of the collision medium. These parameters, identified in the context of $pp$ collisions, serve as benchmarks or references for heavy-ion collision studies. The elevated temperatures and enhanced transverse flow observed in high multiplicities may indicate the potential formation of QGP-like matter, a state of deconfined quarks and gluons. By establishing a connection between freeze-out parameters and system properties in $pp$ collisions, the study contributes to a broader understanding of the dynamics and properties of the interacting medium in heavy-ion collisions, thereby aiding in the exploration of QGP and the broader phases of nuclear matter under extreme conditions.

The relationship between \( q-1 \) and multiplicity class for both 7 and 13 TeV energies is explored in Fig. 2(c). A positive correlation is evident between \( q-1 \) and the multiplicity class, with values at 13 TeV exceeding those at 7 TeV, which is consistent with Refs. \cite{Yang:2022fcj, Li:2022gxy}. The analysis revealed distinct non-equilibrium behaviors for these two particle types. $q-1$ was employed to quantify the deviation from thermal equilibrium, with $q_M$ representing mesons and $q_B$ representing baryons. The parameters $q_M$ and $q_B$ quantify the non-extensivity of mesons and baryons, respectively. Our findings indicate that, within their respective groups, mesons exhibit larger $q$ values compared to baryons, ($q_M$ $>$ $q_B$ ). This distinction suggests that the non-equilibrium characteristics vary between mesons and baryons in the medium formed during pp collisions. This study underscores the importance of considering separate non-extensivity parameters for different particle types, contributing to a more accurate understanding of the non-equilibrium dynamics in the collision system. The filled and empty symbols in fig. 2(c) represent 7 TeV and 13 TeV respectively, while the circle and star symbols show $q_B$ and $q_M$ respectively. $q_M$ is observed to be larger than $q_B$ at a given energy. It is consistent with the conclusions obtained in the case of $pp$ interactions at  low energies \cite{42}. This might make sense in the following way. The hard scattering dominates the production of particles at high $p_T$. The hadron $p_T$ distributions that are similar to the Tsallis distributions are produced by hard scattering of partons  in $pp$ collisions, as demonstrated in Refs. \cite{7, Wong:2015mba}. In addition, the hard scattering hadron spectra exhibit a power law distribution $p_T^{-n}$. The exponent $n$=$1/(q-1)$ relates the index $n$ and $q$, and according to \cite{7, Wong:2015mba, Khandai:2013gva}, $n$ can be written as $n$=$2n_a-4$, where $n_a$ is the quantity of participating active quarks. The counting rule will result in $n = 4$ if parton-parton hard scatterings $qq$ $\rightarrow$ $qq$ (also known as the leading twist process) are the primary processes for hadron production. If the higher twist processes contribution is considered, the index $n$ will grow. For instance, $n$ is equal to 8 in the meson interaction when $q$ + meson $\rightarrow$  $q$ + meson, while $n$ is equal to 12 in the baryon scattering process when $q$ + baryon $\rightarrow$ $q$ + baryon. This results in $q_M$ $>$ $q_B$ by default. It is also observed that compared to $q_B$, $q_M$ grows more with increasing energy. This could be a result of the resonance particles frequently being mesons, which can be important in high-energy collisions. Given that mesons interact with the medium differently than baryons do, the increased production of resonances at higher energies may have contributed to the observed increase in $q$ for mesons.  

The positive correlation between $q$ and the multiplicity class may suggest that, as the system complexity (indicated by the multiplicity class) increases, the non-extensivity of the system also rises. The higher values at 13 TeV could suggest that the system is more out of thermal equilibrium at higher energies \cite{Chen:2020zuw}.
It is worth emphasizing that in the fitting procedure, both the \( \beta_T \) and parameter \( T \) are parameters common to all hadrons under consideration. Conversely, the degrees of non-equilibrium are distinct for mesons (\( q_M \)) and baryons (\( q_B \)), exhibiting universality within their respective groups. This statement suggests that the meson and baryon degrees of non-equilibrium within their respective groups are global. "Universality" in this sense refers to the observation of consistent patterns or behaviors associated with non-equilibrium conditions within each group of particles (baryons and mesons). Despite their differences, mesons and baryons show various degrees of non-equilibrium behavior. All particle types follow this pattern, suggesting a degree of universality among them. Our analysis reveals that employing a single \( q \) value for all particle types yields sub-optimal fits to the experimental spectra. However, when mesons and baryons are treated with separate \( q \) values, the fits significantly improve. It should also be noted that the masses and normalization factors exhibit variability across different particle species.

\begin{figure}
\begin{center}
\includegraphics[width=8.cm]{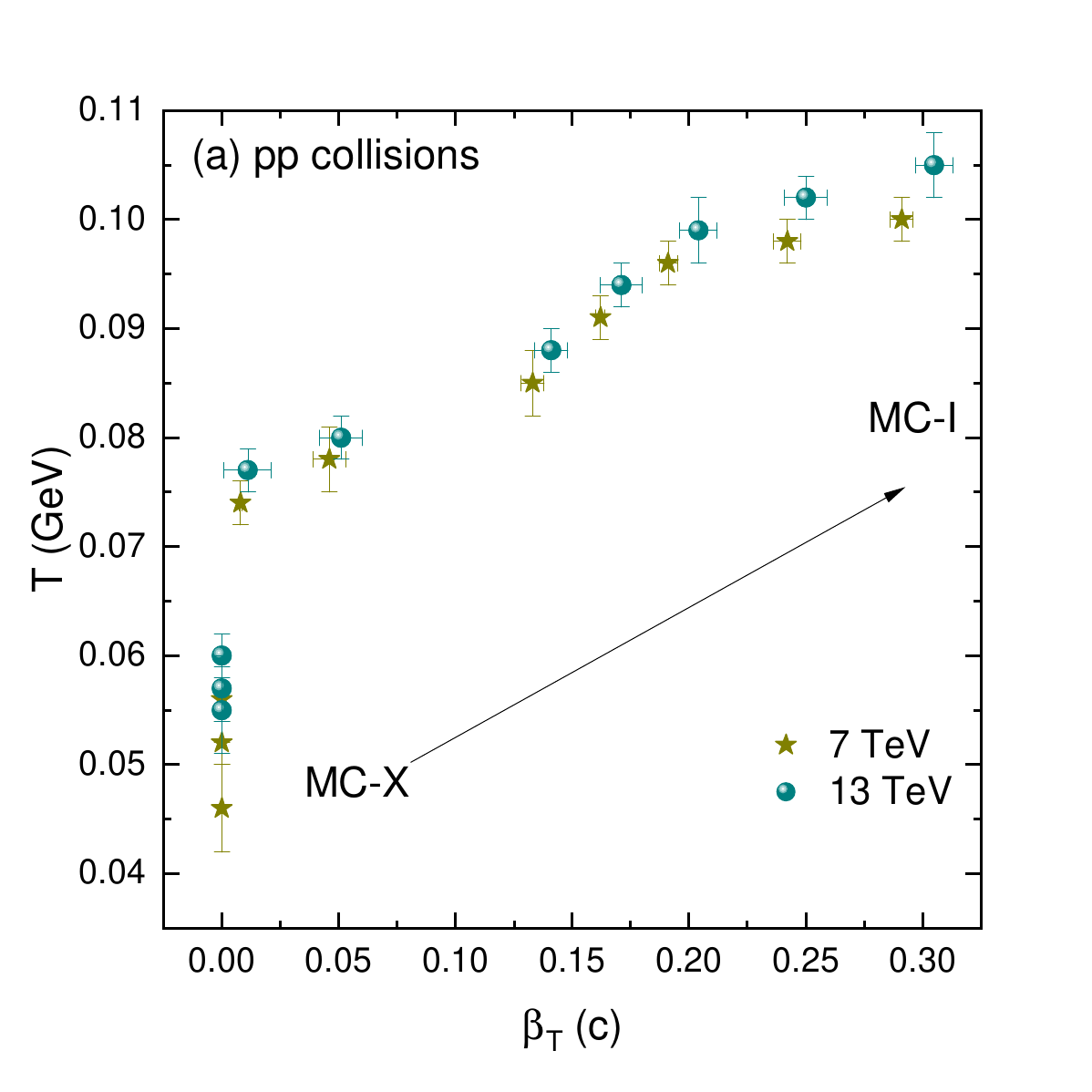}
\includegraphics[width=8.cm]{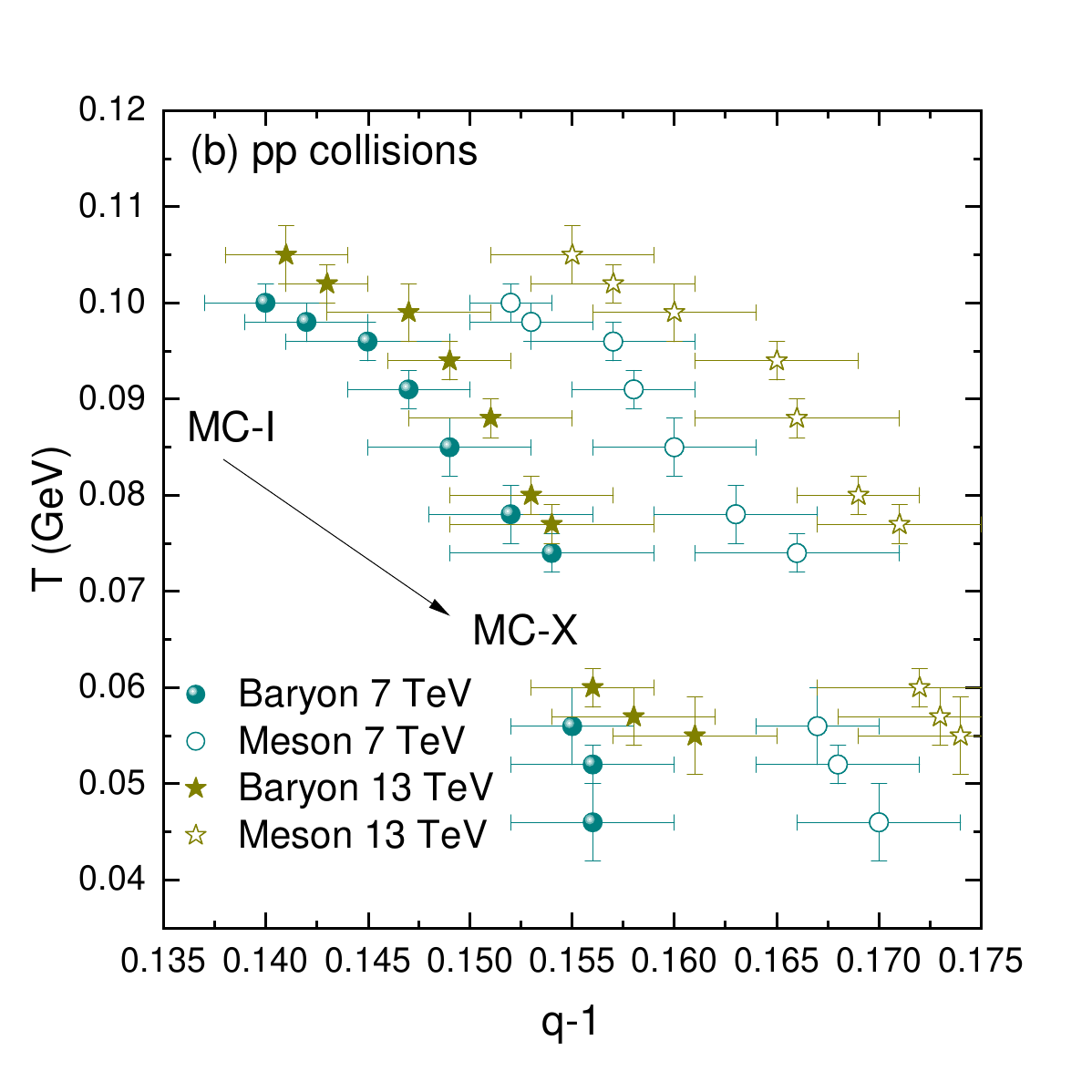}
\caption{This figure displays correlations of $T$ and $\beta_T$, and $T$ and $q-1$ in panels (a) and (b), respectively.}
\end{center}
\end{figure}

\begin{figure}
\begin{center}
\includegraphics[width=18.cm]{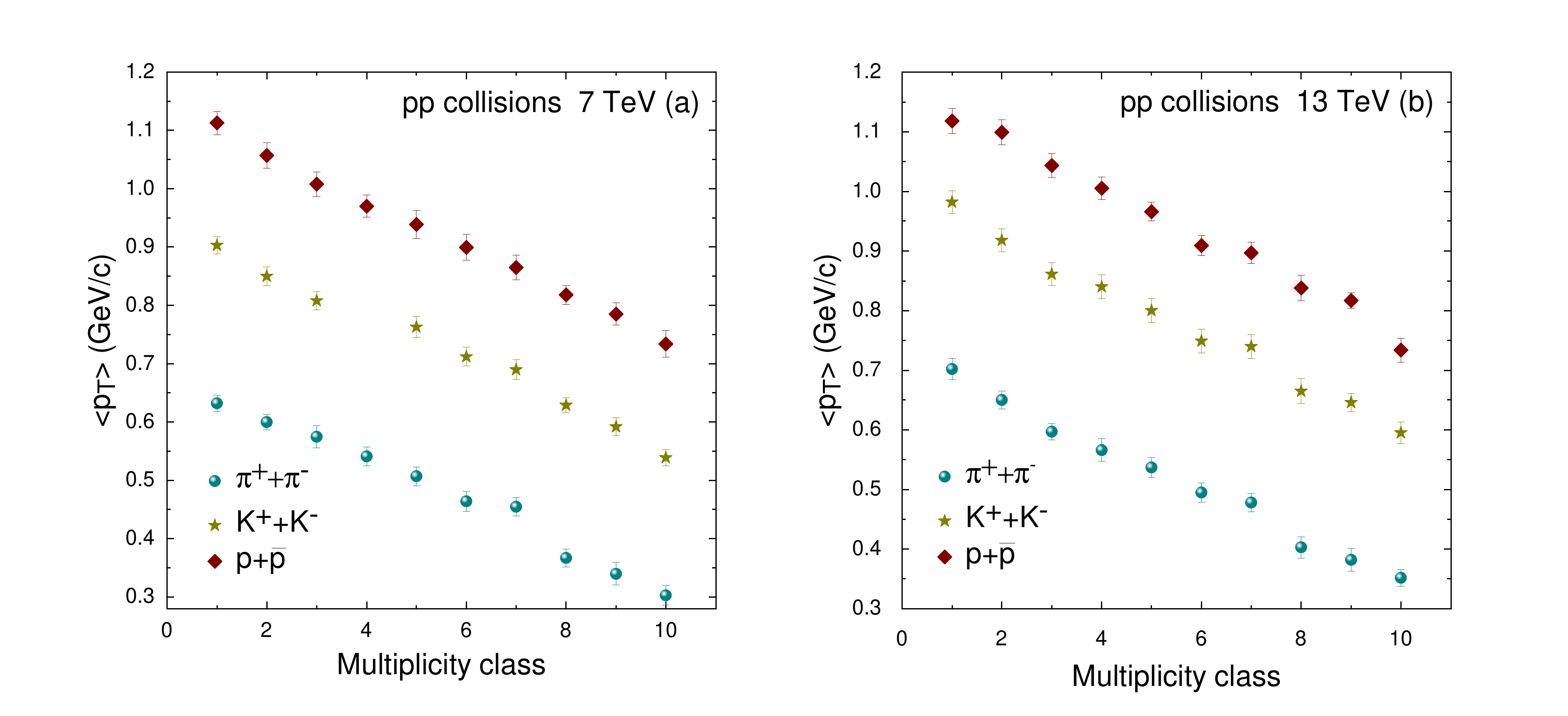}
\caption{Figure 4 illustrates the \( \langle p_T \rangle \) values for the hadrons across different multiplicity classes at energies of (a) 7 TeV (a) and (b) 13 TeV.} 
\end{center}
\end{figure}\begin{figure}
\begin{center}
\includegraphics[width=18.cm]{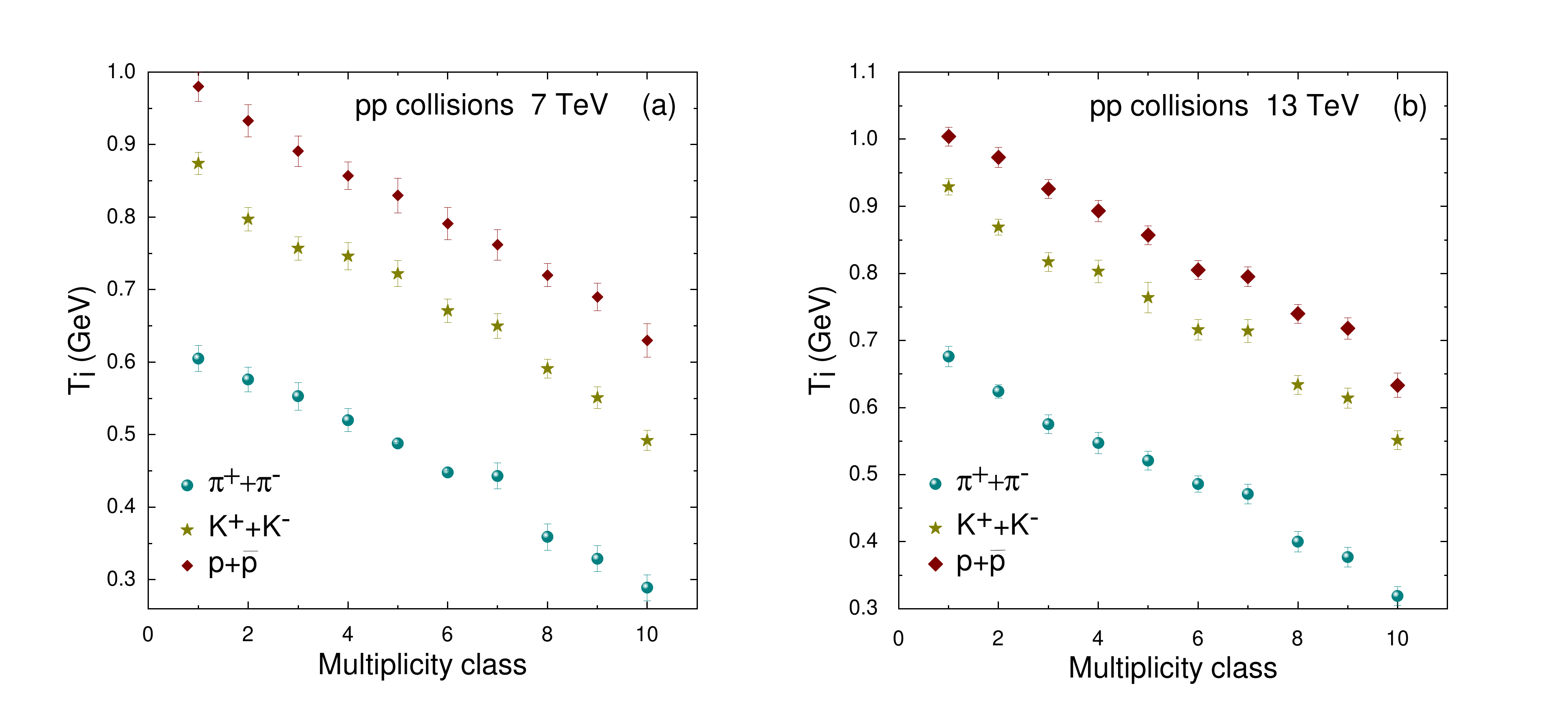}
\caption{The interplay between the initial temperature \( T_i \) and multiplicity class, segmented by particle type are shown here. Panel 5(a) is focused on 7 TeV collisions, while 5(b) presents data at 13 TeV.}
\end{center}
\end{figure}

Figure 3 illustrates the correlations between (a) \( T \) and \( \beta_T \) and (b) \( T \) and \( q-1 \). A positive correlation between \( T \) and \( \beta_T \) is observed in Fig. 3(a). It is found that the evolutions of $T$ and $\beta_T$ at both collision energies follow a similar trend, that is $T$ grows with $\beta_T$ which is influenced by various underlying physical processes. For instance, as the system moves toward thermal equilibrium, both \( T \) and \( \beta_T \) might naturally increase, signifying a more collectively moving and hotter system. Higher values of \( T \) could also reflect greater energy deposition in the collision event, where there is a large pressure gradient which in turn could be manifest as an increase in \( \beta_T \). The initial conditions of the colliding particles, such as their energy and momenta, could likewise impact both \( T \) and \( \beta_T \). The formation of a QGP could influence these parameters in a correlated manner. The \( \beta_T \) and kinetic freeze out temperature \( T_0 \) can be impacted by the formation of QGP by changing the dynamics of the collision process. Depending on the high-energy conditions necessary for the formation of QGP, different patterns of particle interactions and energy transfer can occur during a collision. This, in turn, affects the temperature at which strong interactions between particles "freeze out." A measurement of the collective motion of the particles in the collision, the \( \beta_T \), can also be influenced by the presence and characteristics of the QGP. Quark Gluon Plasma formation essentially modifies the collision dynamics by adding parameters that influence the \( \beta_T \) and consequently the freeze out temperature. 

The relationship might also be affected by event multiplicity, where higher multiplicity often corresponds to more central collisions, resulting in elevated values of both \( T \) and \( \beta_T \) \cite{mult}. However, beyond the multiplicity class VII, there is a sudden fall in $T$ where $\beta_T$ is 0. This can be explained by the collision's initial conditions, which can be extremely important in lower multiplicity events. If the initial energy density is high, even with low multiplicity, it may produce a marginally higher temperature. It is possible that the total amount of energy available for particle production is what is causing this temperature increase rather than collective motion (\( \beta_T \)). Furthermore, processes other than hydrodynamic ones, like particle decays and resonance decays, might dominate low multiplicity events. 

Independent of the transverse flow, these processes may have an impact on temperature measurements. In Fig. 3(b), a negative correlation between \( T \) and \( q-1 \) is observed, and it signifies an inverse relationship between the thermal parameters and the degree of non-equilibrium in the system. This might indicate that as the degree of thermal excitation of the system becomes larger (larger \( T \)), it deviates less from equilibrium (\( q-1 \) decreases), or vice versa. The observed trend where \( T \) decreases and \( q-1 \) increases from lower to higher multiplicity classes (from MC-I to MC-X) could be indicative of a transition from a more thermalized, collective state to a more fragmented, less equilibrated state.
 This could be relevant for understanding the underlying dynamics of particle production and interactions in different multiplicity environments.

Figure 4 elucidates the variations in \( \langle p_T \rangle \) for the considered hadrons as a function of event multiplicity, presented at two collision energies: (a) 7 TeV and (b) 13 TeV. The symbols vertically from top to bottom and horizontally from left to right represent the dependence of \( \langle p_T \rangle \) on particle and multiplicity, respectively. Intriguingly, \( \langle p_T \rangle \) exhibits a decline as one transitions from low to high multiplicity classes, which corresponds to larger \( \langle p_T \rangle \) when multiplicity is larger. Additionally, heavier particles consistently exhibit higher \( \langle p_T \rangle \) values than their lighter counterparts, demonstrating that the radial flow effect rises with particle mass. Notably, the \( \langle p_T \rangle \) values are slightly elevated at 13 TeV relative to 7 TeV, potentially reflecting a little higher available energy in the collision system, as well as a slightly larger radial flow. 

Figure 5 is similar to Fig. 4, but it portrays the relationship between initial temperature \( T_i \) and multiplicity class for all the of hadrons at the two energies. The figure elucidates a negative correlation, showing \( T_i \) to decrease with increasing multiplicity class. This is because a large number of particles is produced in the lower multiplicity class where the initial energy density is higher and results in larger $T_i$ in the lower multiplicity class. Moreover, \( T_i \) values are found to be higher for heavier particles, possibly reflecting their stronger interaction with the medium or higher degrees of freedom. The slightly elevated \( T_i \) at 13 TeV relative to 7 TeV may indicate a little higher thermal or collective effects enabled by the increased collision energy. The trend of $T_i$ and \( \langle p_T \rangle \) is similar, which shows that the latter is the reflection of the former.  

\begin{figure}
\begin{center}
\includegraphics[width=18.cm]{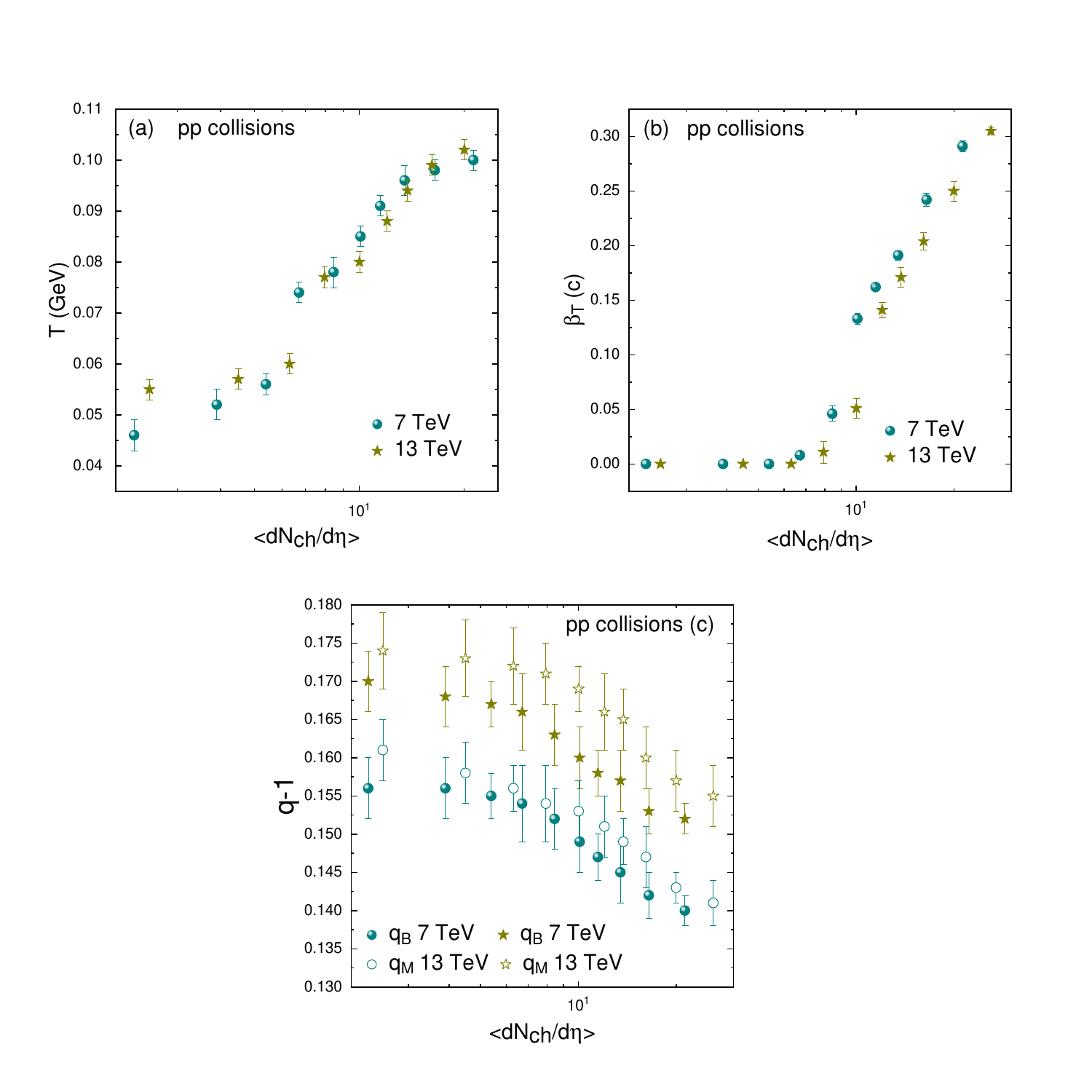}
\caption{This figure comprises three sub-figures exploring the interplay between key parameters and charged particle multiplicity per unit pseudorapidity \( \langle dN_{ch}/d\eta \rangle \) at 7 TeV and 13 TeV. Sub-figures 6(a), and 6(b) show a positive correlation between \( T \) and \( \langle dN_{ch}/d\eta \rangle \), and \( \beta_T \) and \( \langle dN_{ch}/d\eta \rangle \), respectively, while 6(c) shows a negative correlation between \( q-1 \) and \( \langle dN_{ch}/d\eta \rangle \).}
\end{center}
\end{figure}

\begin{figure}
\begin{center}
\includegraphics[width=9.cm]{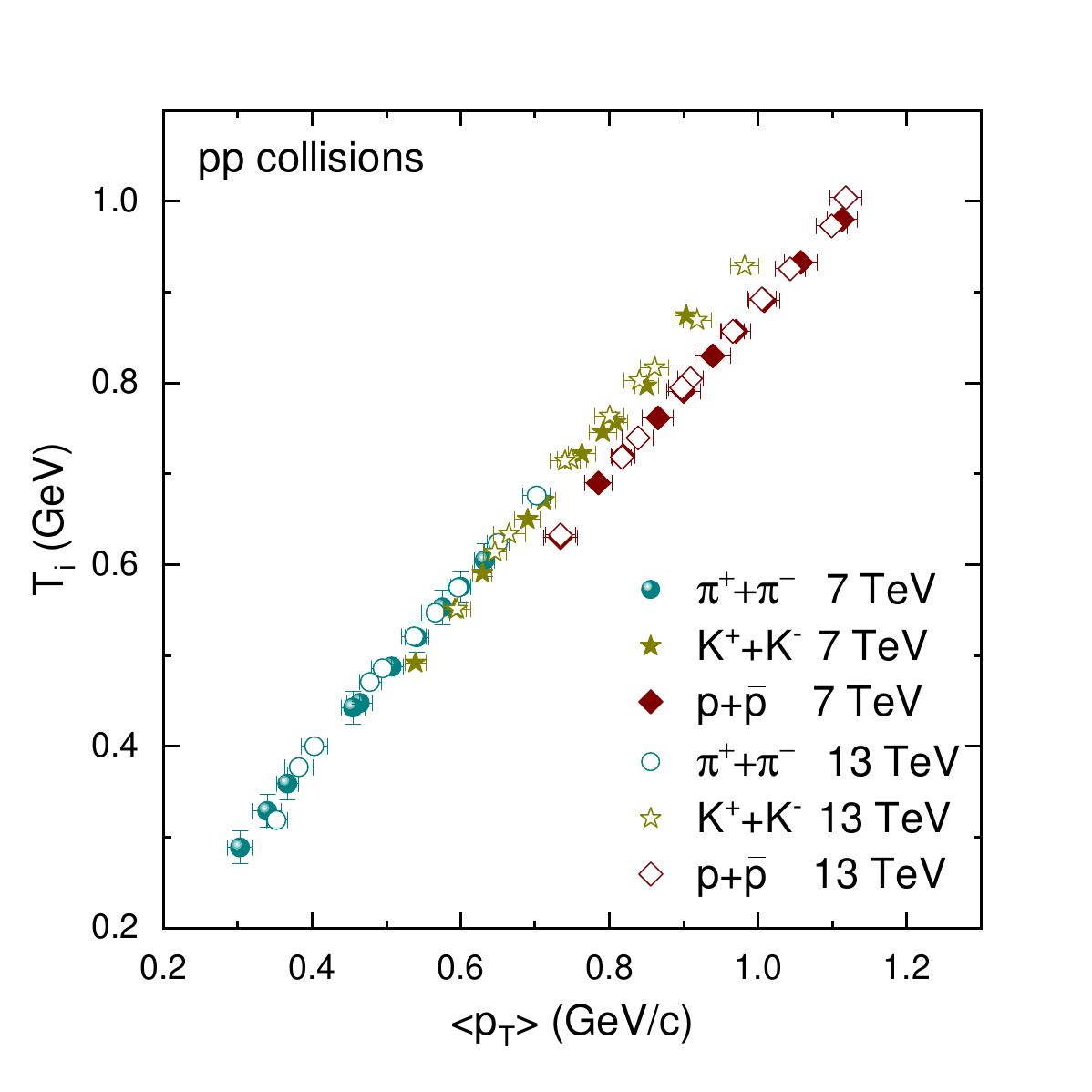}
\caption{Figure 7 explores the relationship between \( T_i \) and \( \langle p_T \rangle \) for the hadrons. The filled symbols show the particles at 7 TeV while the empty symbols represent the particles at 13 TeV.}
\end{center}
\end{figure}

\begin{figure}
\begin{center}
\includegraphics[width=9.cm]{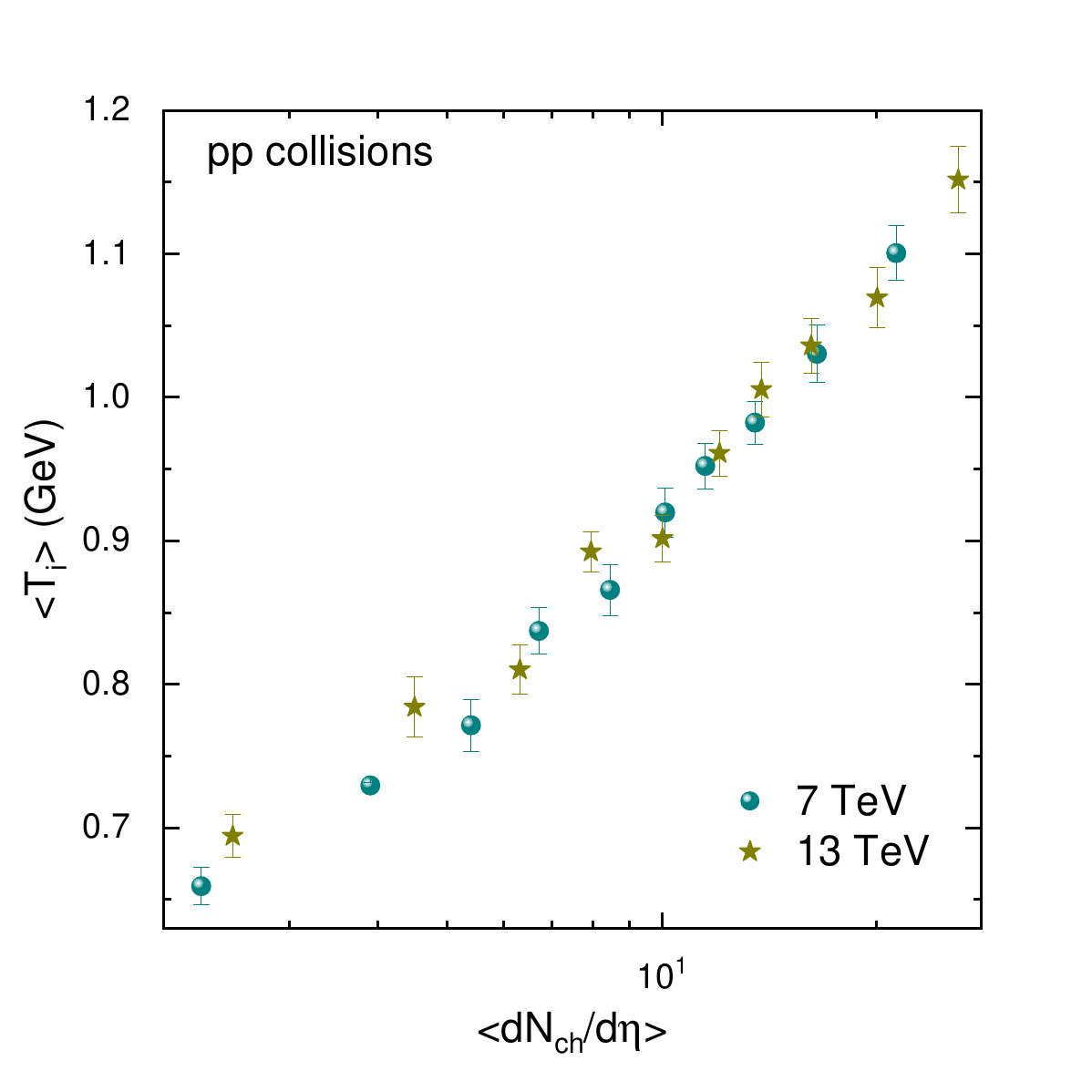}
\caption{This figure presents the correlation between initial temperature \( \langle T_i \rangle \) and charged particle multiplicity per unit pseudorapidity \( \langle dN_{ch}/d\eta \rangle \).}
\end{center}
\end{figure}

\begin{figure}
\begin{center}
\includegraphics[width=9.cm]{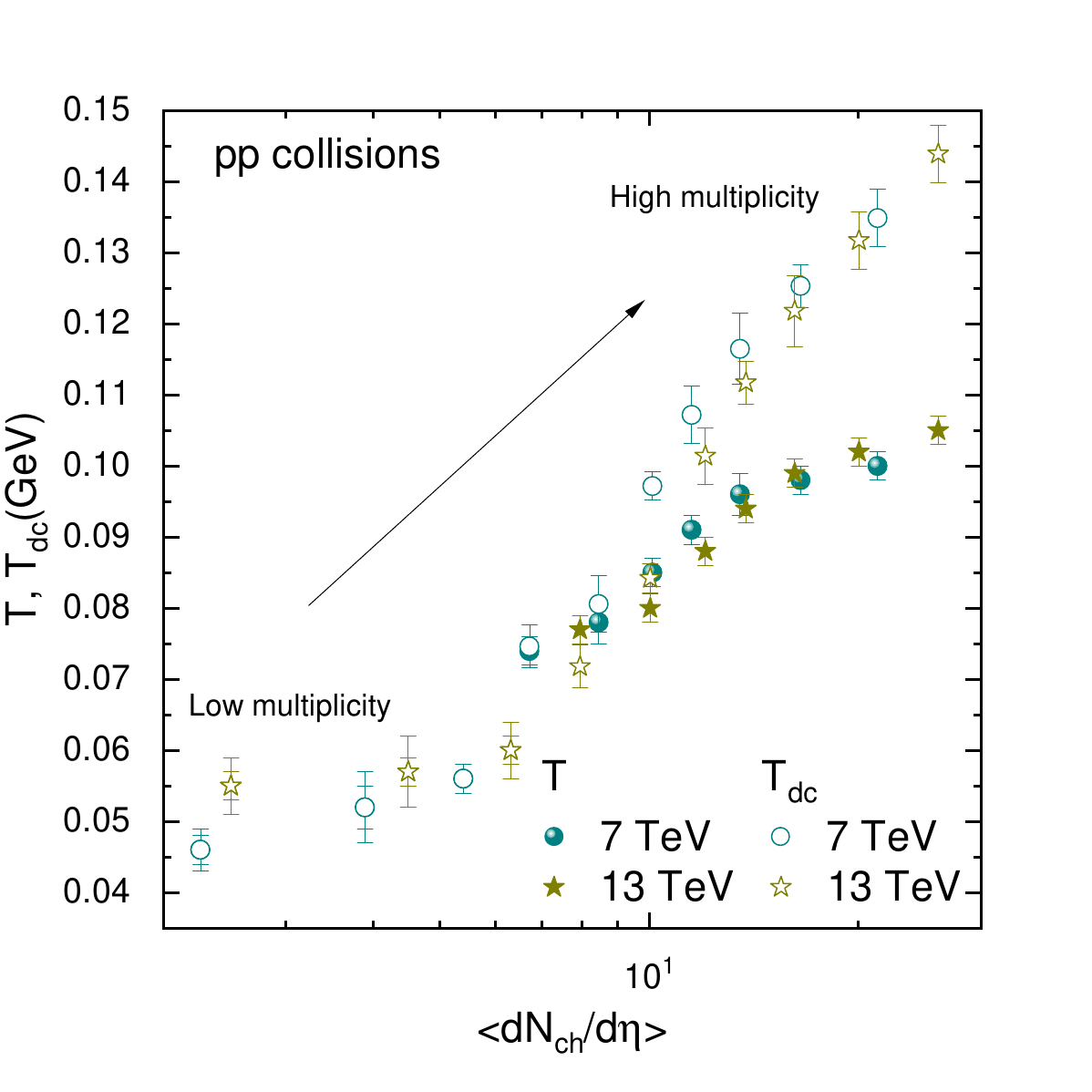}
\caption{ This figure shows the dependence of Doppler-corrected temperature $T_{dc}$ on \( \langle dN_{ch}/d\eta \rangle \)},
and its comparison with the $(T)$.\\
\end{center}
\end{figure}

\begin{figure}
\begin{center}
\includegraphics[width=14.cm]{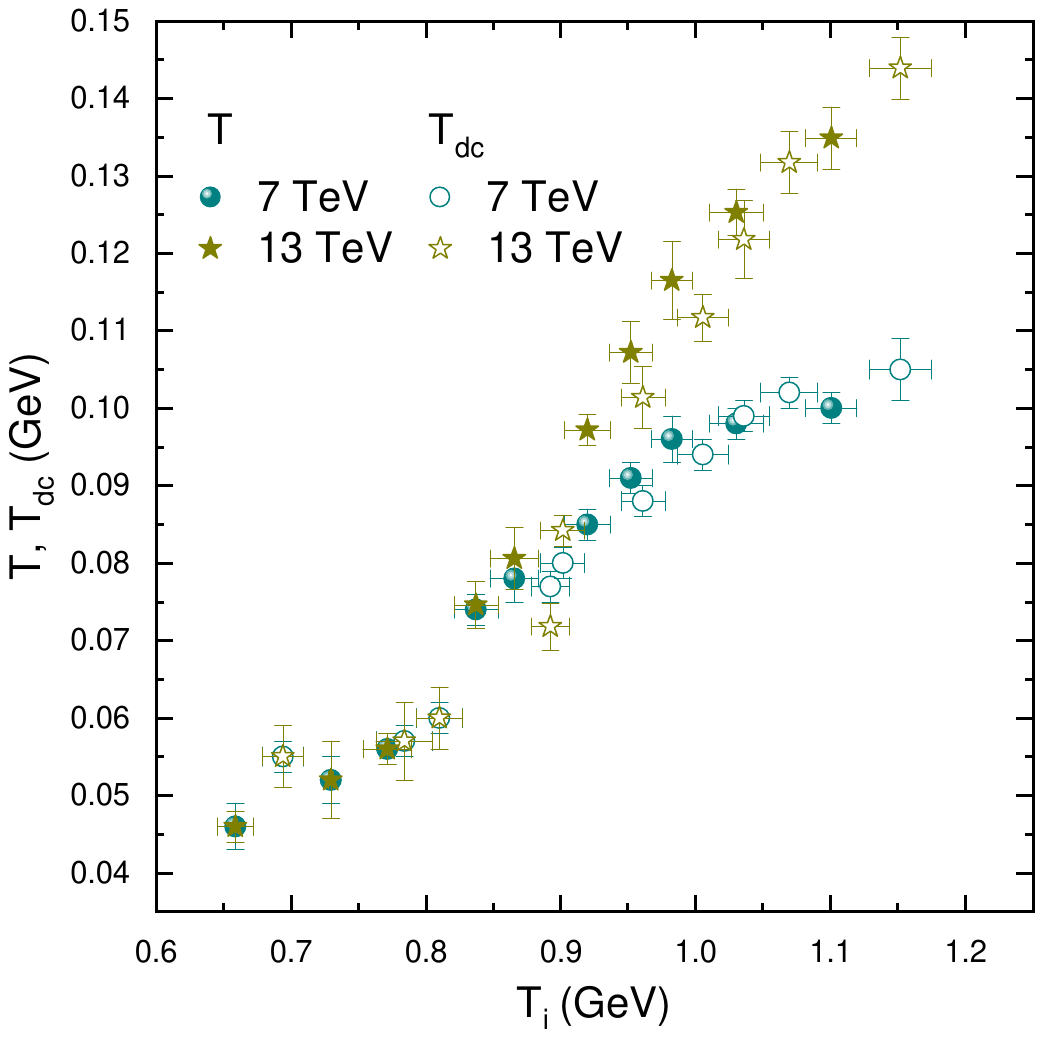}
\caption{ This figure shows the dependence of Doppler-corrected temperature $T_{dc}$ on $T_i$, and its comparison with the $(T)$.}
\end{center}
\end{figure}

Figure 6 comprises three sub-figures, where sub-figure 6(a) elaborates on the correlation between the parameter \( T \) and charged particle multiplicity per unit pseudorapidity \( \langle dN_{ch}/d\eta \rangle \). A positive correlation is observed, suggesting that higher \( \langle dN_{ch}/d\eta \rangle \) values may correspond to increased thermalization within the collision system. The parameters are also higher at 13 TeV compared to 7 TeV, further supporting the notion that elevated collision energies may lead to enhanced thermal and collective effects. Similarly, Fig. 6(b) specifically depicts the relationship between \( \beta_T \) and \( \langle dN_{ch}/d\eta \rangle \) at 7 TeV and 13 TeV. A positive correlation between \( \beta_T \) and \( \langle dN_{ch}/d\eta \rangle \) is evident, suggesting that higher particle densities may drive stronger collective flow. Interestingly, \( \beta_T \) is null for \( \langle dN_{ch}/d\eta \rangle \) smaller than 6, possibly indicating a threshold density below which collective effects are negligible. As in earlier findings, the values are elevated at 13 TeV relative to 7 TeV, potentially reflecting the greater available energy for collective behavior and particle production.

Sub-figure 6 (c) presents the relationship between \( q-1 \) and \( \langle dN_{ch}/d\eta \rangle \), revealing a negative correlation. This suggests that higher particle densities (higher multiplicities) may result in systems that are closer to equilibrium, as evidenced by lower \( q-1 \) values. This can be explained as follows. Compared to low multiplicities, high multiplicities allow the energy fluctuation at initial impact to be completely eliminated by subsequent hadronic interactions. As a result, the former's effect has a smaller impact on the particle spectra than the latter. Additionally, both \( q-1 \) and \( \langle dN_{ch}/d\eta \rangle \) are higher for mesons compared to baryons. 

Figure 7 illustrates the correlation between \( T_i \) and \( \langle p_T \rangle \) for the hadrons at 7 TeV and 13 TeV. A positive correlation is observed, suggesting that higher initial temperatures are associated with greater mean transverse momenta, as discussed above. The positive correlation of $T_i$ and \( \langle p_T \rangle \) is natural because $T_i$ is given by $\sqrt{\langle p^2_T \rangle/2}$. $p+\bar p$ in the higher multiplicity classes can be seen to overlap $K^++K^-$ in lower multiplicity classes, and similarly, the latter in lower multiplicity classes overlap $\pi^++\pi^-$ in higher multiplicity classes. This may indicate that the thermodynamic properties (such as thermalization and hydrodynamic flow) of $\pi^++\pi^-$ in lower multiplicity classes are approximately similar to $K^++K^-$ in higher multiplicity classes, and similarly, the thermodynamic properties of $K^++K^-$ in lower multiplicity classes is approximately similar to $p+\bar p$ in higher multiplicity classes.  

Figure 8 determines the relationship between the  initial temperature  \( \langle T_i \rangle \) and \( \langle dN_{ch}/d\eta \rangle \). A positive correlation is evident, reinforcing the connection between thermal characteristics and particle density. The present study shows that at higher multiplicities, both $T_i$ and \( \langle dN_{ch}/d\eta \rangle \) rise, which characterizes the QGP matter. 

At the point of hadron freezeout, the Doppler-corrected temperature parameter \(T_{dc}\) surpasses the original \(T\) due to a blue shift factor \((T_{dc} = \sqrt{\frac{1+ \langle \beta_T \rangle}{1-\langle \beta_T \rangle}})\) induced by radial flow. Figure 9 illustrates the reliance of the Doppler-corrected temperature on the TBW model. The empty symbols show the Doppler-corrected temperature, while the filled symbols show the \(T\) which we take from Fig. 2(a), in order to compare $T_{dc}$ with $T$. One can see that the Doppler-corrected temperature has an increasing trend with multiplicity, and it shows a stronger dependence on the multiplicity than \(T\). Furthermore, we observe that the correlation among the charged particle multiplicity per unit pseudorapidity \( \langle dN_{ch}/d\eta \rangle \) and $T$ or $T_{dc}$ is weaker at low multiplicities, whereas it becomes stronger towards higher multiplicity. This is because, as the multiplicity grows, their interactions become more significant, and the system takes longer to reach the thermal freezeout stage. It has also been observed that both $T_{dc}$ and $T$ are the same at lower multiplicities (up to the value of 8), after that they separate, and this separation increases as the multiplicity further increases. This can be explained by the fact that the collision system is less dense and the number of produced particles is relatively low at lower multiplicities. The Doppler correction for relativistic motion may not be that important in this regime, so $T_{dc}$ is comparable to $T$. In contrast, more particles are created in the collision as the multiplicity rises. Frequent consequences are higher average velocities and a wider distribution of particle momenta. Because more particles are moving at significant fractions of the speed of light in such circumstances, the Doppler correction becomes increasingly essential. This might result in a circumstance where $T_{dc}$ exceeds $T$. Figure 10 is presented to show the behavior of $T$ and $T_{dc}$ with respect to $T_i$, and is similar to Fig. 9. A linear behavior of $T$ and $T_{dc}$ with $T_i$ is observed. Like Fig. 9, $T$ and $T_{dc}$ are the same at lower initial temperature, and the difference between them increases with the increase of initial temperature.

Before going to the next section, we would like to point out that our study of $pp$ collisions at energies of 7 TeV and 13 TeV suggests intriguing indications of QCD medium formation. Notably, lower multiplicity classes exhibit higher $T$ and $\beta_T$, implying improved thermalization and enhanced collective motion in higher multiplicity scenarios, characteristic of a QCD-like medium. The kinetic freeze-out parameters in our study, particularly the observed higher $T$ and enhanced $\beta_T$ in higher multiplicity events, provide valuable insights into the possibility of QGP-like matter formation, especially in high multiplicity scenarios. The elevated temperatures suggest a more thermally equilibrated system, resembling conditions conducive to QGP formation. The increased transverse flow is indicative of stronger collective motion, a characteristic feature of the QGP, where quarks and gluons exhibit fluid-like behavior. The positive correlation between $q-1$ and multiplicity class indicates an increasing non-equilibrium behavior with system complexity, with higher values at 13 TeV suggesting a more pronounced departure from thermal equilibrium. Treating mesons and baryons with separate $q$ values for better fits suggest distinct non-equilibrium characteristics for different particle types, pointing towards specific features in QCD medium formation. Additionally, the positive correlation between initial temperature $T_i$ and \( \langle p_T \rangle \) reinforces the connection between the initial state and subsequent particle dynamics. The increasing trend of $T_{dc}$ with multiplicity and its separation from $T$ at higher multiplicities highlight the importance of relativistic motion in later collision stages, aligning with expectations for QCD medium presence. Overall, our findings offer valuable insights into the nature of the collision medium in the context of Quantum Chromodynamics. 

The positive correlation between $(q-1)$ and multiplicity class further supports the notion of a more non-equilibrium state with higher system complexity, aligning with expectations for QGP signatures. The distinct non-equilibrium behavior of mesons and baryons, as evidenced by separate $q$ values for better fits, suggests specific characteristics in the formation of QGP-like matter for different particle types. Overall, the observed kinetic freeze-out parameters in high multiplicity events provide indications of QGP-like medium formation in proton-proton collisions. Furthermore, in the case of particle production in proton-lead ($p$+Pb) collisions, we believe that it presents a unique opportunity to study the properties of the Quark-Gluon Plasma (QGP) and the underlying mechanisms of particle production in small collision systems. While $p$+Pb collisions are typically considered reference systems for studying heavy-ion collisions, they also exhibit interesting collective behavior and multiplicity-dependent phenomena similar to those observed in larger collision systems. The study of particle production in p+Pb collisions allows us to explore the interplay between initial-state effects, such as nuclear shadowing and parton saturation, and final-state interactions, such as collective flow and hadronization, in a smaller and more controllable system. By analyzing the multiplicity dependence of various observables, such as particle yields, transverse momentum spectra, and correlations, we can gain insights into the nature of the QGP-like medium formed in these collisions and its evolution with increasing collision centrality or multiplicity. Besides, comparisons between $p$+Pb and $pp$ collisions provide valuable information about the role of nuclear matter effects in particle production and the possible formation of a QGP-like medium in small collision systems.

\section{Summary and Conclusion}

In this study, we have conducted a comprehensive analysis of the freezeout parameters of identified particles (\( \pi^{\pm} \), \( K^{\pm} \), and \( p (\bar{p}) \)) in proton-proton (\( pp \)) collisions at \( \sqrt{s} = 7 \) TeV and \( 13 \) TeV. Utilizing the data provided by the ALICE Collaboration and theoretical framework that combines the TBW, we investigated the \( p_T \) spectra across different multiplicity classes. We used common temperature ($T$) as well as flow velocity ($\beta_T$) for all the particles, but separate non-extensive parameters ($q$), with which the fit results are better. Our fitting parameters, including \( T \), \( \beta_T \), and \( q \), align well with experimental data, confirming the quality of the model.

We found an inverse relationship of the Tsallis temperature (\( T \)) and of the flow velocity ($\beta_T$) with the multiplicity class, in contrast to a linear relationship of the non-extensive parameter ($q$) with multiplicity class. The inverse relation of multiplicity with $T$, and its linear relation with $q$, indicate a quick thermalization in lower multiplicity classes. On the other hand, the inverse relationship of $\beta_T$ with the multiplicity class shows that the system expands monotonously in higher multiplicity classes compared to lower multiplicity classes. $\beta_T$ falls down to zero after the multiplicity class-VII which may show a possible change in the underlying processes, possibly from a regime where the collective effects dominate to one where other factors do.  

Intriguingly, the \( \langle p_T \rangle \) and the $T_i$ values display a decline as we transition from low to high multiplicity classes, while heavier particles consistently show higher \( \langle p_T \rangle \) and $T_i$ values than their lighter counterparts. These observations reflect the thermal and collective behavior of the colliding systems at different multiplicities.

Our results support the notion of different degrees of thermalization and equilibration processes at varying multiplicities. The study also reveals that the degree of non-equilibrium differs between mesons and baryons. This suggests that a universal \( q \) value cannot be applied to all particle types for optimal fit quality. In addition, some useful correlations among the parameters are presented in this study, which provide a better understanding of matter's characteristics in the extreme conditions.

Our findings offer valuable insights into the complex dynamics governing particle production in varying collision scenarios. These results are important for enhancing our understanding of the underlying physics of particle interactions in different environments, contributing to a robust framework for future investigations in high-energy physics.

\section*{Acknowledgment}
This work is supported by the Hubei University of Automotive Technology Doctoral Research Fund under Grant Number BK202313, UNAM-DGAPA through the PAPIIT project IG100322, Ajman University Internal Research Grant No. [DRGS Ref. 2023-IRG-HBS-13]. The authors also extend their appreciation to the Deanship of Scientific Research at Northern Border University, Rafha, Saudi Arabia for funding this research work through the project number "NBU-FFR-2024-2225-04". Finally, the authors express their gratitude to the ALICE Collaboration for allowing the public access to this data.

\section*{Author Contributions} Each listed author has provided significant, direct, and intellectual contributions to the work and has given approval for its publication.

\section*{Data Availability Statement} The data supporting the findings of this study are incorporated within the article and appropriately referenced in relevant sections of the text.

\section*{Compliance with Ethical Standards}

\section*{Ethical Approval} The authors affirm their adherence to ethical standards concerning the content presented in this paper.

\section*{Disclosure} The funding agencies played no part in the study's design, data collection, analysis, interpretation, manuscript preparation, or the decision to publish these findings.

\section*{Conflict of Interest} The authors declare that there are no conflicts of interest regarding the publication of this paper.

\end{document}